\documentclass[nofootinbib,aps,prl,a4paper,twocolumn,superscriptaddress,longbibliography,reprint]{revtex4-2}
\usepackage[english]{babel}
\usepackage[mathlines]{lineno}
\usepackage{amsthm}
\usepackage{amsmath}
\usepackage{amssymb}
\usepackage{dsfont}
\usepackage{graphicx}
\usepackage{bm}
\usepackage{color}
\usepackage[dvipsnames]{xcolor}
\usepackage{enumitem}
\usepackage{mathrsfs}
\usepackage{verbatim}
\usepackage{bbold}
\usepackage{braket}
\usepackage{lipsum}
\usepackage{multirow}
\usepackage{url}
\usepackage{tikz}
\usepackage{graphicx}
\usepackage[colorlinks=true,citecolor=blue,urlcolor=blue]{hyperref}
\usepackage[capitalise,nameinlink]{cleveref}

\newcommand{\cl}[1]{\hat{\mathcal{#1}}}

\usepackage[normalem]{ulem}

\begin{document}   
%\title{Neural network modeling of super- and sub-radiant light-matter-coupled dynamics}
\title{Neural network modeling of many-body super- and sub-radiant dynamics}
\author{Gianluca Lagnese}
\affiliation{Jo\v{z}ef Stefan Institute, 1000 Ljubljana, Slovenia}
\author{Laurin Brunner}
\affiliation{Theoretical Physics III, Center for Electronic Correlations and Magnetism, Institute of Physics, University of Augsburg, D-86135 Augsburg, Germany}
\author{Lorenzo Rossi}
\affiliation{ICFO–Institut de Ciencies Fotoniques, The Barcelona Institute of Science and Technology, 08860 Castelldefels (Barcelona), Spain}
\author{Darrick Chang}
\affiliation{ICFO–Institut de Ciencies Fotoniques, The Barcelona Institute of Science and Technology, 08860 Castelldefels (Barcelona), Spain}
\affiliation{ICREA–Instituci\'{o} Catalana de Recerca i Estudis Avan\c{c}ats, 08015 Barcelona, Spain}
\author{Markus Schmitt}
\affiliation{Forschungszentrum J\"ulich GmbH, Peter Gr\"unberg Institute 8 (Quantum Control), 52425 J\"ulich, Germany}
\affiliation{Institute of Theoretical Physics, University of Regensburg, D-93053 Regensburg, Germany}
\author{Zala Lenar\v{c}i\v{c}}
\affiliation{Jo\v{z}ef Stefan Institute, 1000 Ljubljana, Slovenia}

\begin{abstract} 

There is significant interest in exploring novel phenomena in quantum light-matter interfaces, which are driven by the combination of structured dissipation and long-range interactions that are typical in such systems. To this end, it is important to develop new general numerical simulation techniques, which can access large system sizes and are not based on semi-classical approaches. 
%In most experimental realizations, light-matter coupling induces all-to-all interactions, amenable to simple mean-field descriptions, with physics closer to a semi-classical than to a genuinely many-body quantum one. Setups that can evade the mean-field description include subradiant ordered arrays interacting via dipolar transitions, with structured, suppressed light emission. Due to a structured dissipation and quasi-long-range interactions, they are significantly more challenging to model numerically.
Here, we report the first application of neural quantum states to obtain the dissipative dynamics of light-matter-coupled systems beyond what is accessible with exact and tensor-network calculations.
We specifically apply this method to simulate the many-body emission dynamics of approximately 40 atoms, arranged in dense arrays in one and two dimensions. These systems have been chosen because they can support prominent subradiant dynamics at late times and could be realized with cold atomic quantum simulators.
%which are empirically difficult to capture by semi-classical techniques.
%Modeling about 40 atoms in one and two dimensions, we reveal the many-body character of subradiant dynamics and put in direct comparison different geometries. Our approach unlocks the possibility of studying many-body phases that could be realized with quantum simulators of densely packed arrays.
\end{abstract}

\maketitle
% \begin{itemize}
%     \item Novel development in simulators with light-matter interaction: 
%     \begin{itemize}
%         \item combination of short-range inherent and long-range light-mediated interactions
%         \item ordered arrays with interactions and dissipation due to interference effects
%     \end{itemize}
%     \item Traditional methods are not capable of covering above combinations of short and long-range interactions, as well as dissipation.
%      \begin{itemize}
%         \item Review the status of currently available methods for light-matter interacting problems
%         \item TWA, cumulants expansion, TN, DMFT
%         \item what do they fail to cover
%     \end{itemize}   
%     \item Here we propose NN approach. Overview, how this is different to previous studies, how we advance the field [markus, laurin]
% \end{itemize}

While the phenomenon of superradiance was first proposed by Dicke decades ago~\cite{dicke_coherence_1954}, collective emission dynamics in many-body systems of highly excited atoms continues to produce surprising behavior~\cite{gonzalez-tudela_efficient_2017,guerin_subradiance_2016,henriet_critical_2019,holzinger_beyond_2025,pineiro_orioli_emergent_2022,pineiro_orioli_subradiance_2020,masson_universality_2022,lee_exact_2025,qu_spin_2019,Olmos25}. To gain further insights, it is imperative to develop reliable numerical techniques that go beyond the limited system sizes accessible by exact simulations. To this end, semi-classical techniques, such as mean-field or cumulant methods, and discrete truncated Wigner approximation have been widely employed and empirically work well to capture the emission dynamics at early times and from initial product states~\cite{qu_spin_2019,henriet_critical_2019,mink_collective_2023,robicheaux_beyond_2021,bigorda2023,Hosseinabadi25}. However, they generally fail at later times~\cite{bigorda2023,robicheaux_beyond_2021,henriet_critical_2019,mink_collective_2023}, when the system enters into subradiant regimes characterized by the slowdown of decay processes. This suggests that subradiance might support more genuine quantum many-body effects, and necessitates the development of alternative, general numerical methods. 

% \zala{[Add a section on SOTA of ML for open systems ]}
In recent years, artificial neural networks (ANNs) have emerged as a new tool for compressed quantum state representations, that can be exploited for efficient numerical simulations.
Nascent methods based on neural quantum states (NQS) have been shown to extend computational capabilities to address challenging regimes of quantum many-body physics in and out of equilibrium \cite{Carleo2017,Schmitt2020,Schmitt2022,Nys2024,Chen2024,Djuric2025,Roth2025,Gu2025}.
The methodical developments include new approaches to investigate dynamics and steady states of open quantum systems \cite{Hartmann2019,Vicentini2019,Reh2021,Luo2022,Vovk2025}.
However, especially in view of dynamics, previous work mostly focused on the exploration of different methodical avenues and the application to benchmark problems, where the revealed physical phenomenology remained limited.

In the following, we employ the time-dependent variational principle for the probabilistic NQS representation as introduced in Ref.~\cite{Reh2021}. We perform a high-precision study revealing sub-radiance in the dynamics of one- and two-dimensional ordered, densely packed arrays of cold atoms, which experience photon-mediated long-range dipolar interactions and correlated dissipation.
The use of NQS opens access to modeling truly many-body aspects of subradiance, previously not accessible due to small system sizes amenable to exact and tensor network simulations. Having an approach that can address variable geometries allowed us to confirm the existence and thermodynamic stability of a many-body subradiant regime in one and two-dimensional arrays. By considering different architectures, we explore their efficiency to represent different regimes - from super-radiant to sub-radiant - in a protocol most suitable for experimental realization.  
%\textcolor{orange}{Give a few more details and summary of findings? (new architecture, system sizes beyond ED/TN, establish, that sub-radiant regime is not affected by finite size)}

{\it Model.} 
We study one- and two-dimensional ordered atomic arrays in free space, interacting via photon emission and absorption in a dipole-allowed transition between two levels separated by an energy $\hbar\omega_0$. We consider a regime in which the light field can be integrated out, leading to an effective spin description of the problem. In the rotating frame with respect to $\omega_0$, the Liouvillian describing the light-mediated interactions and structured dissipation reads \cite{Asenjo-Garcia17,gross1982,agarwal1970}
\begin{align}\label{eq:lindblad}
\cl{L}\rho &= -\frac{i}{\hbar}\left[H , \rho\right]
+ \sum_{i,j} \Gamma^{ij} \left( \sigma^-_j \rho \sigma^+_i - \frac{1}{2} \{ \sigma^+_i \sigma^-_j,\rho \} \right),\notag\\
H &= \hbar\sum_{i,j}J^{ij} \sigma^+_i \sigma^-_j,
\end{align}
with exchange coupling $J^{ij}$ and the decay rate $\Gamma^{ij}$ given by the tensor $\textbf{G}(\textbf{r}_i, \textbf{r}_j, \omega_0)$ corresponding to the  Green's function of the classical electromagnetic wave equation at frequency of the transition $\omega_0$,
\begin{align}\label{eq:JandGamma}
J^{ij} &= -\frac{\mu_0 \omega_0^2}{\hbar} \, \textbf{p}_i^\dagger \cdot \textrm{Re} \textbf{G}(\textbf{r}_i, \textbf{r}_j, \omega_0) \cdot \textbf{p}_j, \\
\Gamma^{ij} &= \frac{2\mu_0 \omega_0^2}{\hbar} \, \textbf{p}_i^\dagger \cdot \textrm{Im} \textbf{G}(\textbf{r}_i, \textbf{r}_j, \omega_0) \cdot \textbf{p}_j.\notag
\end{align}
Here, $\mu_0$ is the vacuum permeability and $\textbf{p}_j$ is the dipole matrix element of atom $j$.  

For tightly trapped atoms with essentially fixed positions, we will use the free-space Green's tensor  $\textbf{G}_0(\textbf{r}_{ij}, \omega_0) \equiv \textbf{G}(\textbf{r}_i, \textbf{r}_j, \omega_0)$ with $\textbf{r}_{ij} = \textbf{r}_i - \textbf{r}_j$. Explicitly, 
\begin{align}\label{eq:green}
\textbf{G}_0(\textbf{r}, \omega_0) 
= \frac{e^{i k_0 r}}{4 \pi k_0^2 r^3} [ &(k_0^2 r^2 + i k_0 r -1)\mathbb{1}\notag \\
&+ (3 - 3ik_0 r -k_0^2 r^2) \frac{\textbf{r}\otimes \textbf{r}}{r^2}]
\end{align}
with $r=|\textbf{r}|$ and $k_0 = 2\pi / \lambda_0=\omega_0/c$ \cite{Asenjo-Garcia17}.
We will consider an ordered atomic array with inter-atom distance $d$.
The natural time unit is set by the decay rate of a single isolated atom, 
$\Gamma_0 = \Gamma_{jj} = \omega_0^3 |\textbf{p}|^2 / 3\pi \hbar \epsilon_0 c^3$, where $c$ is the speed of light and $\epsilon_0$ is the vacuum permittivity.  As is clear from \cref{eq:green}, the positions of atoms $\textbf{r}_j$ play a crucial role by structuring the interaction and the dissipation through interference effect of the light medium. We will consider open boundary array that are natural to realize with cold atom platforms.

At the level of single magnon excitations, the physical implications of \cref{eq:lindblad,eq:JandGamma,eq:green} are well-understood~\cite{Asenjo-Garcia17}. A magnon excitation whose wavevector is larger than $k_0$ cannot radiatively decay due to momentum mismatch with propagating photons. Such perfectly subradiant solutions exist only if $k_0$ lies within the first Brillouin zone, which requires an interatomic spacing to be $d<\lambda_0/2$ in 1D and $d<\sqrt{2} \lambda_0$ in 2D. On the other and, magnons with wavevectors smaller than $k_0$ will generally decay. It has already been pointed out that if the system is initialized in a highly excited many-body state, one observes a two-stage dynamics with a fast transient super-radiant decay of excitation \cite{Masson22}, corresponding to a burst of photons out of the system, followed by a slow, critical-like power-law decay of excitations in the subradiant regime~\cite{Loic19}. While the physics of the transient super-radiant regime and its relation to the Dicke super-radiance is rather well understood~\cite{Masson22}, much less is known about the subradiant regime, apart from the non-interacting, single-excitation arguments given above. 
%Part of the reason lies in the up-to-date numerics; for a microscopic calculation on small system sizes accessible with exact and tensor network calculation, the subradiant regime for a simple to prepare initial product state appears at rather small density of excitations, close to the non-interacting regime \cite{Loic19}.
Part of the reason is that the subradiant regime generally emerges at a low density of excitations \cite{Loic19}. However, on small systems sizes accessible with exact and tensor network calculations, the low-density and single-excitation regimes become synonymous. 

\begin{figure*}
    \centering
    \includegraphics[width=0.8\linewidth]{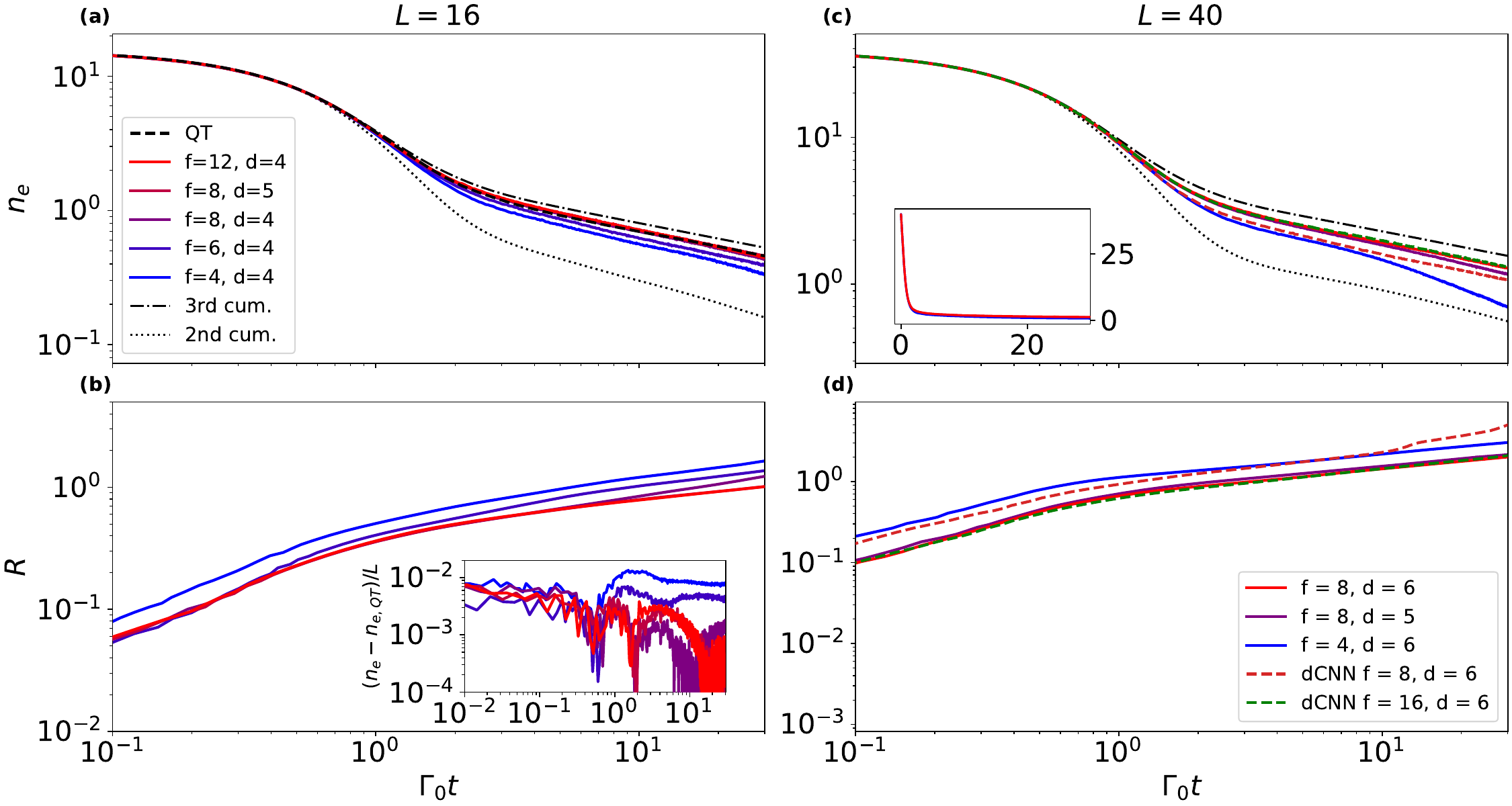}
    \caption{(a) and (b): benchmarking of NQS ResNet results against quantum trajectories (QT) for a linear array of $L=16$ atoms for different architecture parameters and  kernel size $k=4$. In (a) we show the time evolution of the number of excited atoms $n_e$, computed with the NQS and compared with the quantum trajectories prediction (dashed) and contrasted with 2nd (dotted) and 3rd (dot-dashed) cumulant expansion. In (b) we monitor the cumulative error of the TDVP algorithm $R(t)$, \cref{eq:error}.  In the inset of (b) we show the difference between ResNet  and QT results, divided by the system volume $L$. (c) and (d): convergence of time evolution with NQS method by varying network parameters for a linear array of $L=40$ atoms. Here ResNet (solid) is compared with the dCNN (dashed) architecture. The kernel size is fixed to $k=6$ for ResNet and to $k=2$ for dCNN. The NQS results are again contrasted with the 2nd and 3rd order cumulant expansion. In (c) we show the time evolution of the selected observable, in (d) we monitor the accumulated TDVP error. The inset in (c) shows the same data but in linear scale.}
    \label{fig:resnet1d}
\end{figure*}

%\begin{figure}
%    \centering
%    \includegraphics[width=1\linewidth]{convergence1d.pdf}
%    \includegraphics[width=1\linewidth]{L16resnet.pdf}
%    \caption{Benchmarking of Resnet against quantum trajectories for L=16. In (a) we show the time evolution of the observable (number of excited atoms) computed with the NN and contrasted with the quantum trajectories prediction. In the inset of (a) we show the difference between NN and quantum trajectories results. In (b) we monitor the cumulative error of the TDVP algorithm. \textcolor{red}{maybe reference to equation --- should we show a much worse case ??}}
%    \label{fig:resnet1dL16}
%\end{figure}
%
%
%\begin{figure}
%    \centering
%    \includegraphics[width=1\linewidth]{L40resnet.pdf}
%    \caption{Convergence of time evolution with the NN method by varying network parameters.  In (a) we show the time evolution of the selected observable, in (b) we monitor the accumulated TDVP error. \textcolor{red}{we should decide wether to keep it like this or to propagate more, as for L=16}}
%    \label{fig:resnet1dL40}
%\end{figure}
{\it Method.}
In the following, we will employ a variational approach for efficient numerical time evolution.
It is based on the description of the density matrix through the probabilities $P(\mathbf a) = \text{tr}(M^\mathbf{a}\rho)$ of measuring an element of a many-body positive operator-valued measure (POVM).
Following previous work \cite{Carrasquilla2019,Reh2021}, we choose the tetrahedral POVM as the local single-qubit operator basis.
Since this choice is informationally complete (IC-POVM), the density matrix is fully specified by the outcome probabilities $P(\mathbf a)$.
Using the Lindblad equation leads to a master equation $\frac{d}{dt}P(\mathbf a)=\sum_{\mathbf a'}\mathscr L^{\mathbf a\mathbf a'}P(\mathbf a')$ for the probabilities, where the operator $\mathscr L$ is derived from the Liouvillian $\mathcal L$ \cite{Reh2021}.
A more in depth description of this ansatz can be found in the Supplementary Material.

% In the following, we will employ a variational approach for efficient numerical time evolution. It is based on an expansion of the many-body density matrix in terms of a positive operator-valued measure (POVM).
% %
% Each POVM element $M^{\mathbf a}$ is associated with a measurement outcome $\mathbf a$, that is obtained in a state $\rho$ with probability
% \begin{align}
%     P(\mathbf a)=\text{tr}\left(\rho M^{\mathbf a}\right)\ .
%     \label{eq:povm_prob}
% \end{align}
% For a composite Hilbert space $\mathcal H=(\mathcal H_{\text{local}})^{\otimes N}$, the many-body POVM can be constructed via the tensor products of the measurement operators $M_{\text{loc}}^a$ of a local POVM \cite{Carrasquilla2019}: $M^{\mathbf a}=M_{\text{loc}}^{a_1}\otimes\ldots\otimes M_{\text{loc}}^{a_N}$.
% In the following, we will work with the tetrahedral POVM $M_{\text{loc}}^{a}=(\mathds 1+\mathbf{s}^a\cdot\mathbf\sigma)/4$, where $a=0,1,2,3$ and $\mathbf{s}^a$ point to the corners of a tetrahedron on the Bloch sphere.
% %
% This POVM is minimal and informationally complete, meaning that the relation \eqref{eq:povm_prob} can be inverted, yielding
% \begin{align}\label{eq:density_matrix_from_POVM}
%     \rho=\sum_{\mathbf{a},\mathbf a'}P(\mathbf a)\left(T^{-1}\right)^{\mathbf a,\mathbf a'}M^{\mathbf a'}\ ,
% \end{align}
% where $T^{\mathbf a,\mathbf a'}=\text{tr}\left(M^{\mathbf a}M^{\mathbf a'}\right)$ is the overlap matrix.
% %
% Therefore, the state $\rho$ can be fully described by specifying the $4^N$ probabilities $P(\mathbf{a})$.
% %

POVM-based NQS \cite{Carrasquilla2019} use an ANN $P_{\mathbf{\theta}}(\mathbf a)$ with parameters $\mathbf\theta$ as compressed representation of the quantum state.
The universal approximation theorems for ANNs guarantee that any architecture-specific bias eventually vanishes when increasing the network size \cite{Cybenko1989, Hornik1991}, providing a way to systematically improve and assess the accuracy of simulations.
By contrast to other ways of representing a density matrix as an NQS \cite{Hartmann2019,Vicentini2019} POVM-based density matrices are Hermitian irrespective of the ANN architecture; however, the resulting density matrices are not necessarily positive semi-definite. 
%

%
% While there exist alternative approaches to solve the time evolution of the density matrix \cite{Luo2022,Vovk2025}, we employ a time dependent variational principle (TDVP) \cite{Reh2021} to evolve the network parameters $\theta$.
% In first order in the time step $\tau$, the POVM-probability distribution is given by $P_{\theta(t)}(\mathbf{a}) + \tau \mathcal{L}^{\mathbf{a}\mathbf{b}}P_{\theta(t)}(\mathbf{b})$.
% Minimizing the Kullback-Leibler divergence between this and the distribution from updating the network parameters $P_{\theta(t) + \tau \dot\theta}(\mathbf{a})$ up to second order in $\tau$, we get the TDVP equation
% We employ a time dependent variational principle (TDVP) \cite{Reh2021} to evolve the network parameters $\theta$ for an approximate solution of the master equation.
We employ a time dependent variational principle (TDVP) \cite{Reh2021}, which prescribes the optimal evolution of the NQS parameters $\theta$ via a TDVP equation $S_{kk'} \dot\theta_k = F_k$.
% The projection of the master equation onto the tangent space of the manifold defined by $P_\theta(\mathbf a)$% is obtained by minimizing the Kullback-Leibler divergence $\mathcal{D}_\mathrm{KL}\left(P_{\theta(t+\tau)}, \exp(\mathscr L\tau)P_{\theta(t)}\right)$ for a small time step $\tau$, which 
% leads to the TDVP equation $S_{kk'} \dot\theta_k = F_k$.
Here $S_{kk'} = \langle \chi_k \chi_{k'} \rangle-\langle \chi_k\rangle\langle \chi_{k'} \rangle$ is the quantum geometric tensor, where $\chi_k(\mathbf a)=\frac{\partial}{\partial\theta_k} \log P_\theta(\mathbf a)$ denotes the logarithmic derivatives and $\langle\cdot\rangle$ stands for expectation values with respect to $P_\theta(\mathbf a)$. $F_k = \langle \chi_k \mathscr{L} \rangle-\langle \chi_k\rangle\langle \mathscr{L} \rangle$ is called the force vector.
Expectation values are estimated using Monte Carlo sampling. The parameter update is obtained by solving the TDVP equation, which requires a (pseudo-)inverse of $S_{kk'}$ with careful regularization \cite{Schmitt2020,Hofmann2022}. For this purpose, we employ techniques established in previous work \cite{Schmitt2020, jvmc, Medvidovic2023,Nys2024} and we introduce an adaptive variant of the pseudo-inverse truncation; details are provided in the Supplementary Material (SM).
%
%\zala{The accuracy of the time evolution is controlled by several numerical parameters: the tolerance $\text{PinvTol}$ used in the inversion of $S_{kk'}$; the number of Monte Carlo samples $N_{\mathrm{MC}}$; a $\text{SNRTol}$ cutoff controlling the signal-to-noise ratio (SNR) of the force vector due to Monte Carlo noise \cite{Schmitt2020}; and the tolerance of the adaptive integrator, which determines the time step.}

For a discrete time step $\tau$ the Kullback-Leibler (KL) divergence $\mathcal{D}_\mathrm{KL}\left(P_{\theta(t+\tau)}, \exp(\mathscr L\tau)P_{\theta(t)}\right)$ quantifies the deviation between approximate and exact solution.
We use the accumulated KL divergence 
\begin{equation}\label{eq:error}
    R(t) = \int_0^t \mathrm{d}t' \mathcal{D}_\mathrm{KL}\left(P_{\theta(t') + \tau \dot\theta}, P_{\theta(t')} + \tau \mathcal{L}P_{\theta(t')} \right)
\end{equation}
to assert the accuracy of the obtained solutions.

% Expectation values of observables can be calculated straightforwardly using \eqref{eq:density_matrix_from_POVM}
% \begin{equation}
%     \langle O\rangle = \mathrm{tr}(\rho O) = P(\mathbf{a}) \Omega^\mathbf{a},
% \end{equation}
% where $\Omega^\mathbf{a} = \mathrm{tr}\left((T^{-1})^{\mathbf{a},\mathbf{a'}} M^\mathbf{a'} O\right)$ is independent of the state of the system.
% In practice, these expectation values are calculated by Monte Carlo sampling the distribution $P_\theta(\mathbf{a})$.

Within this work we consider two different neural network architectures for the NQS: a ResNet and a dilated convolutional neural network (dCNN). The ResNet architecture follows closely the construction of Ref.~\cite{Chen2024} and its size is controlled by the kernel size $k$, network depth $d$, and the number of features (or channels) $f$. The dCNN architecture, with different biases at finite network sizes, follows the ideas introduced in Ref.~\cite{wavenet2016}. It is based on the application of convolutional filters with a certain degree of dilation, which is exponentially increased with each layer in the depth direction. Both architectures are described in more detail in the End Matter.
%The hyperparameters used are listed in the respective figures or captions.
The hyperparameters controlling the expressivity are listed in corresponding figures.

{\it Results in one-dimension.}
For the 1D case, we consider the model at inter-atom distance $d=\lambda_0/5$ and with polarization of dipoles along the chain, described by \cref{eq:lindblad} with the couplings \cref{eq:JandGamma} obtained from \cref{eq:green} by imposing the condition $\textbf{p} \cdot \textbf{r} = |\textbf{p}| r$.
First we benchmark our NQS algorithm against the quantum trajectories approach (QT) \cite{daley14}. With each trajectory calculated by exact diagonalization, the dynamics of observables is approximated via a mean over sampled trajectories. In \cref{fig:resnet1d}~\textbf{(a)} and \textbf{(b)} we show the dynamics starting from an initial configuration in which all atoms are in their excited state, for an array of $L=16$ atoms and $10^4$ trajectories. Furthermore, we compare our results with  the cumulant expansion approach \cite{kubo1962,kramer2015,robicheaux2021,plankensteiner2022}, an approximate technique assuming factorization of correlations, see SM, thus more appropriate to describe mean field physics. Previous applications of cumulant expansion to problems of correlated emission and superradiance reported reasonable agreement at short times \cite{rubies2023,kumlin25}.
The simulation with neural networks is performed by selecting the ResNet architecture and checking for convergence with increasing network size. In particular, the kernel size $\mathrm{k}=4$ is fixed, while we vary the network depth $d$ and the number of features $f$. 
%\textcolor{red}{In all the simulation we fixed the number of Monte Carlo samples to $N_{\text{MC}} =10^5$, the inversion tolerance $\text{PinvTol} = 10^{-6}$ and the signal-to-noise ratio tolerance to $\text{SNRTol}=2$.}
Additional details on the network architecture and modeling parameters are given in the End Matter.

\begin{figure*}
    \centering
    \includegraphics[width=0.8\linewidth]{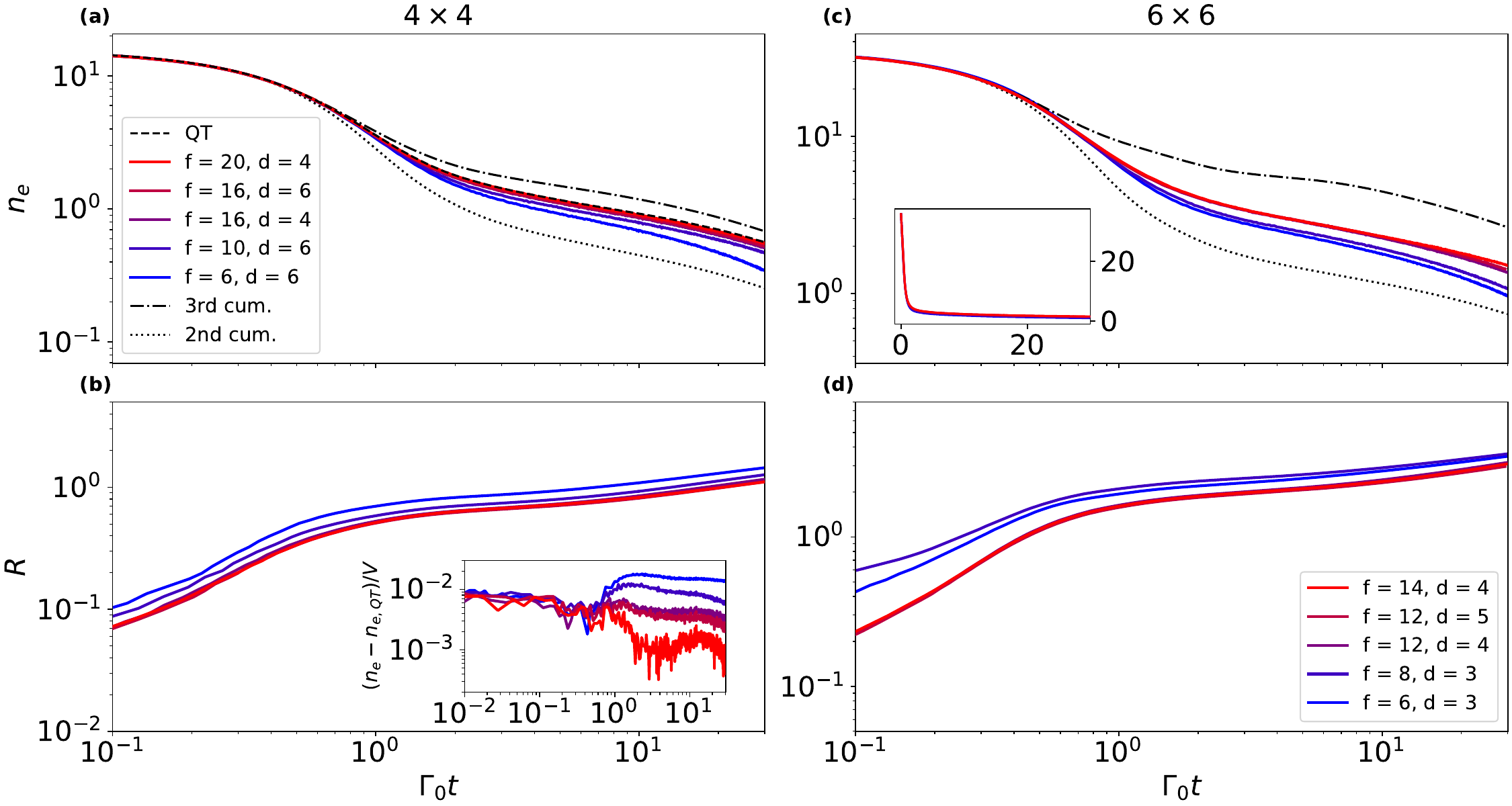}
    \caption{
    (a) and (b): benchmarking of ResNet against quantum trajectories for a square lattice of $V=4\times4$ atoms for different architecture parameters, with kernels size fixed to $2\times2$. In (a) we show the time evolution of the number of excited atoms computed with the NQS compared with the quantum trajectories (black dashed) prediction and contrasted with the 2nd (dotted) 3rd (dot-dashed) cumulant expansion. In the inset of (b) we show the difference between ResNet and quantum trajectories results divided by the number of atoms $V$. In (b) we monitor the cumulative error of the TDVP algorithm $R(t)$.  (c) and (d): convergence of time evolution with the NQS method by varying network parameters for a square lattice of $6\times6$ atoms. The kernel size is fixed to $3\times3$. In (c) we show the time evolution of the selected observable, in (d) we monitor the accumulated TDVP error. The inset in (c) shows the same data but in linear scale.}
    \label{fig:resnet2d}
\end{figure*}

In panel \textbf{(a)} of \cref{fig:resnet1d} we follow the total number of excited atoms, $n_e = 1/2\sum_i  (\sigma^z_i + \mathbb{1}_i )$, as an easily measurable observable that reveals the existence of different dynamical regimes. As reported above and in Refs.~\cite{Asenjo-Garcia17,masson_universality_2022,Loic19}, dynamics consist of initial super-radiant regime for $\Gamma_0 t \lesssim 1$ and later subradiant, emergently critical regime for $\Gamma_0 t \gtrsim 1$ with power-law decay of occupation. In panel \textbf{(b)} we track the cumulative TDVP error, as defined in \cref{eq:error}. Our benchmarking confirms that upon increasing the expressivity of the network by its depth $d$ and feature $f$, we converge to the exact results. The inset in panel \textbf{(b)} shows the difference between the ResNet and QT predictions, normalized by the system size $L$: despite being significantly affected by noise due to Monte Carlo sampling, it provides the order of magnitude of the achieved precision. The cumulant expansions agrees with the other methods at early times, but start to deviate from the exact results in the subradiant regime, underlining that the latter cannot be fully explained in terms of simple mean-field description.

In order to address the thermodynamic stability of subradiant regime and the pertinence of many-body effects, we simulate the same protocol on previously inaccessible $L=40$ system size, shown in \cref{fig:resnet1d}~\textbf{(c)} and \textbf{(d)}. The number of network parameters is progressively increased until convergence is reached, both, for the observable \textbf{(c)} and the cumulative TDVP error in panel \textbf{(d)}. 
In the insets of panel \textbf{(c)} the same data are shown in linear scale. 
%We highlight how the phenomenon of superradiance is not easily detectable on a linear scale and the numerical method we are testing is required to achieve a high level of precision. 
To support convergence, we report also the results obtained with the other architecture, i.e. the dilated CNN (dashed lines). Although only two parameter choices are reported, convergence is achieved with the same approach as for the ResNet. Furthermore we mention that, in order for the two architectures to achieve agreement, the dilated CNN seems to require a smaller number of parameters ($N_\text{par}=2976$) compared to the ResNet ($N_\text{par}=4544$), see the End Matter for further comparison.

On system sizes tractable with exact diagonalization, the onset of subradiant dynamics happens for the maximally excited initial state at about one or two excitations left in the system, implying proximity to the non-interacting regime, see \cref{fig:resnet1d}~\textbf{(a)}. Our results for $L=40$ show that, as the system size increases, the system enters the subradiant regime while multiple excitations are still present, indicating the need for a many-body treatment. The character of the subradiant regime can be further explored through other observables, such as two-point correlators shown in the End Matter, confirming that NQS Ansatz captures correlations to a similar level of precision as the number of excitations $n_e$ presented here.

{\it Results in two-dimensions.}
We now proceed to explore the sub-radiant dynamics in 2D arrays, which have been so far mostly treated with semiclassical methods \cite{Masson22,Wang24, kumlin25}. Here, we consider the orthogonal photon polarization  with couplings \eqref{eq:JandGamma} obtained from \cref{eq:green} at  $\textbf{p} \cdot \textbf{r} = 0$. The results at inter-atom distance $d=\lambda_0/5$ are shown in \cref{fig:resnet2d}. The NQS algorithm is again benchmarked against quantum trajectories and contrasted with 2nd and 3rd order cumulant expansion. The results are provided in panels \textbf{(a)} and \textbf{(b)} for a square lattice of $V=4 \times4$ atoms. 
For the NQS simulation with ResNet architecture, the kernel size is fixed to $k = 2 \times 2$ and number of features $f$ and the depth $d$ are varied. 
%The results appear in the legend with increasing number of parameters from the bottom to the top. The convergence towards the QT results, displayed in \cref{fig:resnet2d}\textbf{(a)}, corresponds to an overall decreased value of the accumulated TDVP error shown in \textbf{(b)}. 
The deviation of NQS results from the QT results in the inset of \textbf{(b)} gives the magnitude of the error. Compared to the 1D case, the deviation of 2nd and 3rd order cumulant expansion from the exact result appears to be larger, once again underlining correlated nature of subradiance. 

\begin{figure}
    \centering
    \includegraphics[width=1\linewidth]{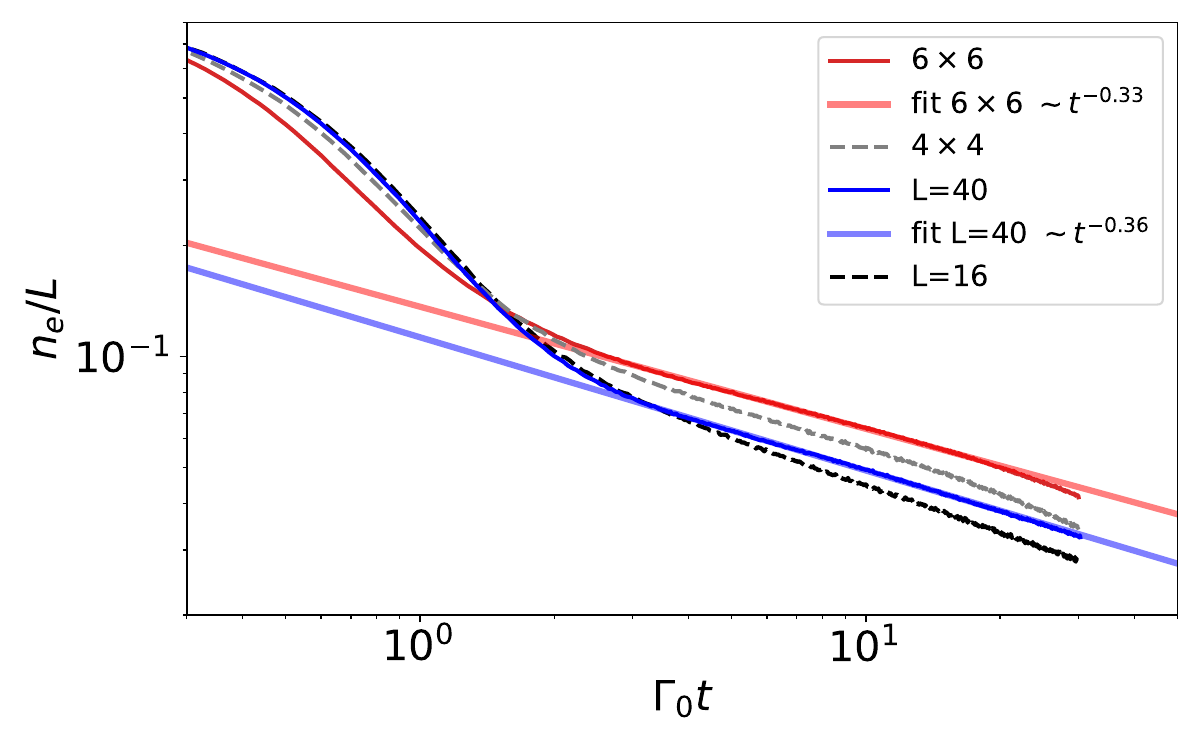}
    \caption{Comparison of dynamics for density of excitation at different system sizes and different geometries. Only exact/best results for the given system size are shown. Late time dynamics correspond to a power law decay, fitted on largest system sizes. }
    \label{fig:densities}
\end{figure}

To explore dynamics at system sizes beyond the reach of alternative methods, we consider an array of $V=6 \times 6$ atoms in panels \textbf{(c)} and \textbf{(d)}. The size of the kernel is fixed to $k = 3 \times 3$, while the depth $d$ and the number of features $f$ increase progressively. 
%
%\textcolor{red}{The number of Monte Carlo samples $N_{\text{MC}}$, the inversion tolerance $\text{PinvTol}$ and the signal-to-noise ratio tolerance to $\text{SNRTol}$ are fixed as for the one dimensional case.}
%
In the legend the results are organized so that the number of network parameters increases from bottom to top. The accumulated TDVP error is shown in panel \textbf{(d)}. Unlike in the 1D case, we do not consider dilated CNNs here. In 2D, the linear system size—which sets the upper bound on correlation length—scales only as $\sqrt{V}$, so dilated convolutions are not expected to provide significant benefits for $\sqrt{V}=6$, though they may become useful at larger scales. Even though there are no available exact data to compare with, the quality of the prediction provided by the cumulant expansion seems to deteriorate. In particular the 3rd order displays a possibly spurious flexions at intermediate times and appears farther away from the NQS results.
Finally, our results confirm that subradiant dynamics in 2D is characteristically similar to 1D, see \cref{fig:densities} and the discussion below.

{\it Discussion and Conclusions.}
We showcase the first-time application of neural quantum states to dissipative light-matter coupled dynamics beyond reach of exact and tensor-network simulation methods. We consider dynamics in sub-wavelength ordered atom arrays, starting from a simple, maximally excited state and displaying various regimes: early time super-radiant regime and late time subradiant regime due to correlated emission, characteristic of ordered sub-wavelength arrays. We show that NQS are versatile enough to capture both regimes and the transition between them. This is a non-trivial result, given that the physical properties and the characteristic correlations in the two regimes are very different.\\
The NQS approach, which can address a range of system sizes and different geometries, manages to put those in concrete comparison, summarized in \cref{fig:densities}. At the level of excitation density $n_e/L$, subradiant dynamics in 1D and 2D appear quite similar, at least for the maximally excited initial state considered: (a) The exponent of the power-law decay characteristic of subradiant emergently critical regime \cite{Loic19} is very similar and close to $1/3$. Reasoning for such exponent and its universality is left for a future study. (b) For this initial state the subradiant dynamics is entered at rather low excitation densities, which increase as the system size is increased, making its character more many-body on thermodynamically large systems. 
%Capturing the subradiant regime numerically is rather challenging: it appears at long times and for easily preparable initial states at rather low excitation densities, for which one needs high accuracy to follow. At smaller system sizes considered up to now, the onset of subradiant dynamics happens at densities that are essentially in the non-interacting regime. Here we considered larger $L=40$ system size in 1D and $6 \times 6$ system size in 2D, confirming that (a) on larger system sizes one can explore the many-body aspects of subradiant regime and (b) that the critical-like power-law decay happens also in 2D geometry. 
\\
Our approach can be used for simulating other, more advanced experimental protocols with cold atom dipole arrays, such as adding interactions and driving protocols, that could stabilize many-body subradiant effects at higher, more easily tractable excitation densities.
Moreover, an alternative realization of physics described here is also in terms of arrays of solid-state excitons \cite{suarez24,Huang25,lagoin24,morin25}.
In general, our study underlines the power of POVM based neural network computing for dissipative dynamics at challenging conditions, such as following critical power-law relaxation on long time scales, simulation of quasi-long range interactions and dissipators, and implementation on various geometries.

\begin{acknowledgments}
Z.L. and G.L. acknowledge the support by the program P1-0044 of the Slovenian Research and Innovation Agency (ARIS), the ERC StG 2022 project DrumS by Horizon Europe, Grant Agreement 101077265, and the European Union Horizon 2020 under the QuantERA II project QuSiED (No 101017733).
Z.L. also acknowledges Markus Heyl and the Guest professorship program of the University of Augsburg, where initial ideas for the implementation of the project have been discussed.
M.S.~was supported through the Helmholtz Initiative and Networking Fund, Grant No. VH-NG-1711.
The NQS simulations were implemented using the jVMC codebase \cite{jvmc}.
\end{acknowledgments}

% Selling points:
% \begin{itemize}
% \item first-time application of NQS to light-matter coupled dynamics beyond reach of exact/TN simulation methods
% \item first many-body calculation of subradient dynamics in one and two dimensions beyond what is reachable with exact diagonalization
% \item confirmation of truly many-body character of subradiant regime; due to small system sizes considered, previous studies observed subradient decay in the regime of single excitation
% \item First results for subradient critical-like decay in 2D.
% \end{itemize}

%\bibliography{biblio,biblio_dc}
\bibliography{biblio}

\section*{END MATTER}

\subsection{Different Architectures: Comparison of ResNet and Dilated CNN}
\begin{figure}[b!]
    \centering
    \includegraphics[width=1.\linewidth]{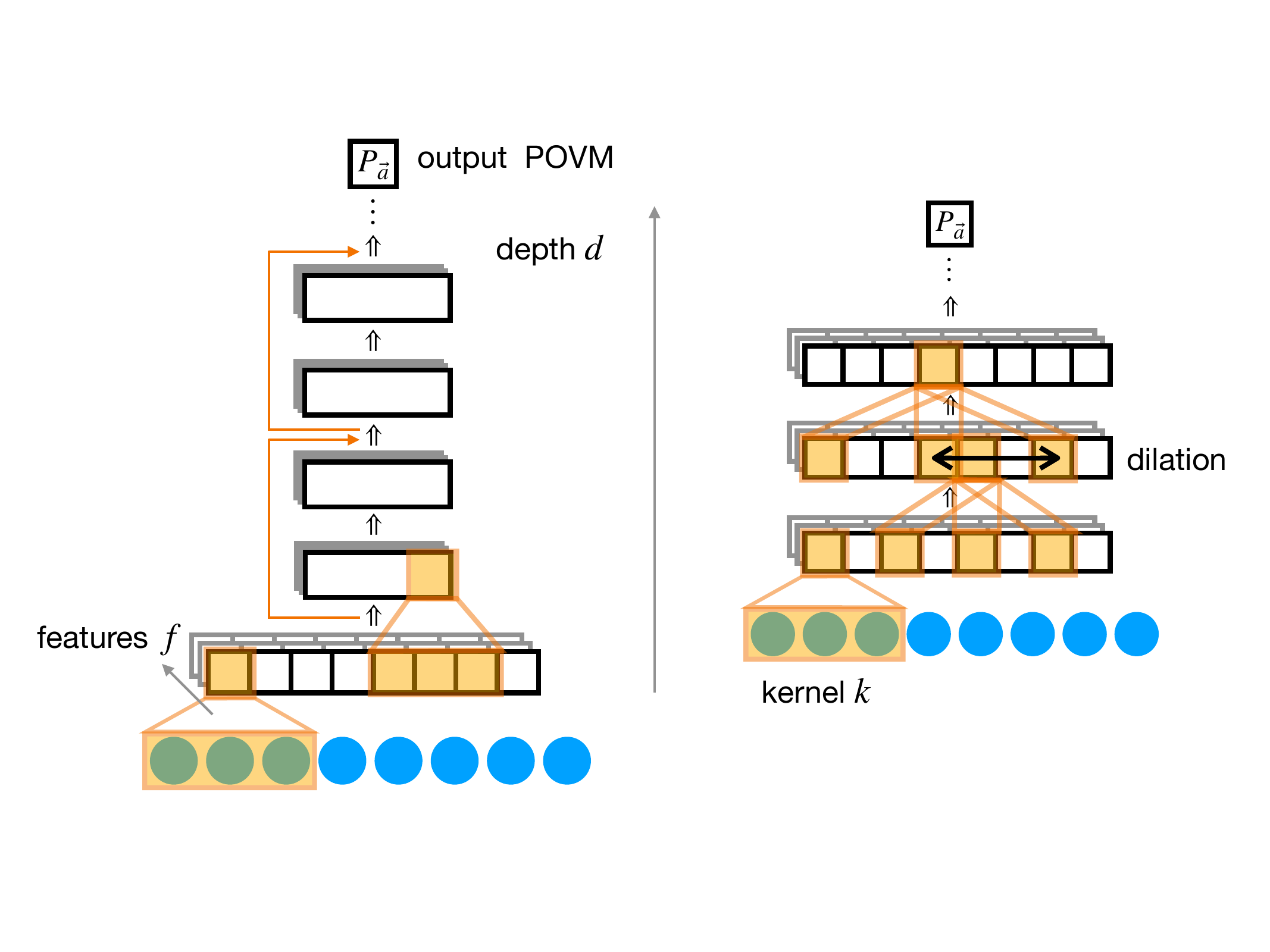}
    \caption{Sketch of the two architectures discussed in the main text. Left: representation of the ResNet architecture. Right: representation of the dilated convolutional neural network (dCNN).}
    \label{fig:sketch}
\end{figure}

In order to capture the correlations generated by the super- and sub-radiant   dynamics, we consider two complementary neural-network architectures implementing distinct strategies. The two approaches are sketched in \cref{fig:sketch}. The relevant quantity is the effective receptive field $RF$, i.e., the spatial range over which input degrees of freedom can influence a given output. A minimal requirement is that it covers the system size, $RF \sim L$. 

In the residual network (ResNet), each layer performs a local operation with kernel size $k$, such that the receptive field grows linearly with depth $d$, $RF \sim k\,d$. The presence of skip connections (orange lines) ensures a well-conditioned parametrization of deep architectures, avoiding pathological gradient suppression and enabling an efficient evaluation of the TDVP equations. This is crucial for accurately propagating the state within the variational manifold.

In contrast, the dilated convolutional neural network (dCNN) enlarges the receptive field at each layer by introducing a dilation factor, leading to a cumulative growth $RF \sim k \sum_{l} r_l$. For exponentially increasing dilation $r_l \sim 2^l$, this yields $RF \sim k\,2^d$, such that distant degrees of freedom are coupled already at shallow depth.

While both approaches can, in principle, encode long-range correlations, the former builds them progressively through depth, whereas the latter incorporates them explicitly through a multiscale connectivity pattern. This comparison allows one to assess whether dynamically generated long-range correlations are more efficiently captured via local compositions between layers or via an explicit enlargement of the receptive field.

In \cref{tab:resvscnn} we compare the two architectures for the simulations shown in \cref{fig:resnet1d}~(c) of the main text in terms of the total number of network parameters. We find that the dCNN achieves comparable performance with a significantly reduced number of parameters.

\begin{table}
\caption{Total number of network parameters for the simulations showed in \cref{fig:resnet1d}(c) of the main text.}
\begin{ruledtabular}
\begin{tabular}{c c c c c}
Architecture & $f$ & $d$ & $k$ & $N_{\mathrm{par}}$ \\
\hline
\multirow{3}{*}{ResNet} 
 & 8 & 6 & 6 & 4544 \\
 & 8 & 5 & 6 & 3760 \\
 & 4 & 6 & 6 & 1216 \\
\hline
\multirow{2}{*}{dCNN} 
 & 16 & 6 & 2 & 2976 \\
 & 8  & 6 & 2 & 848  \\
\end{tabular}
\end{ruledtabular}
\label{tab:resvscnn}
\end{table}

\subsection{Other observables}

\begin{figure}[t]
    \centering
    \includegraphics[width=1\linewidth]{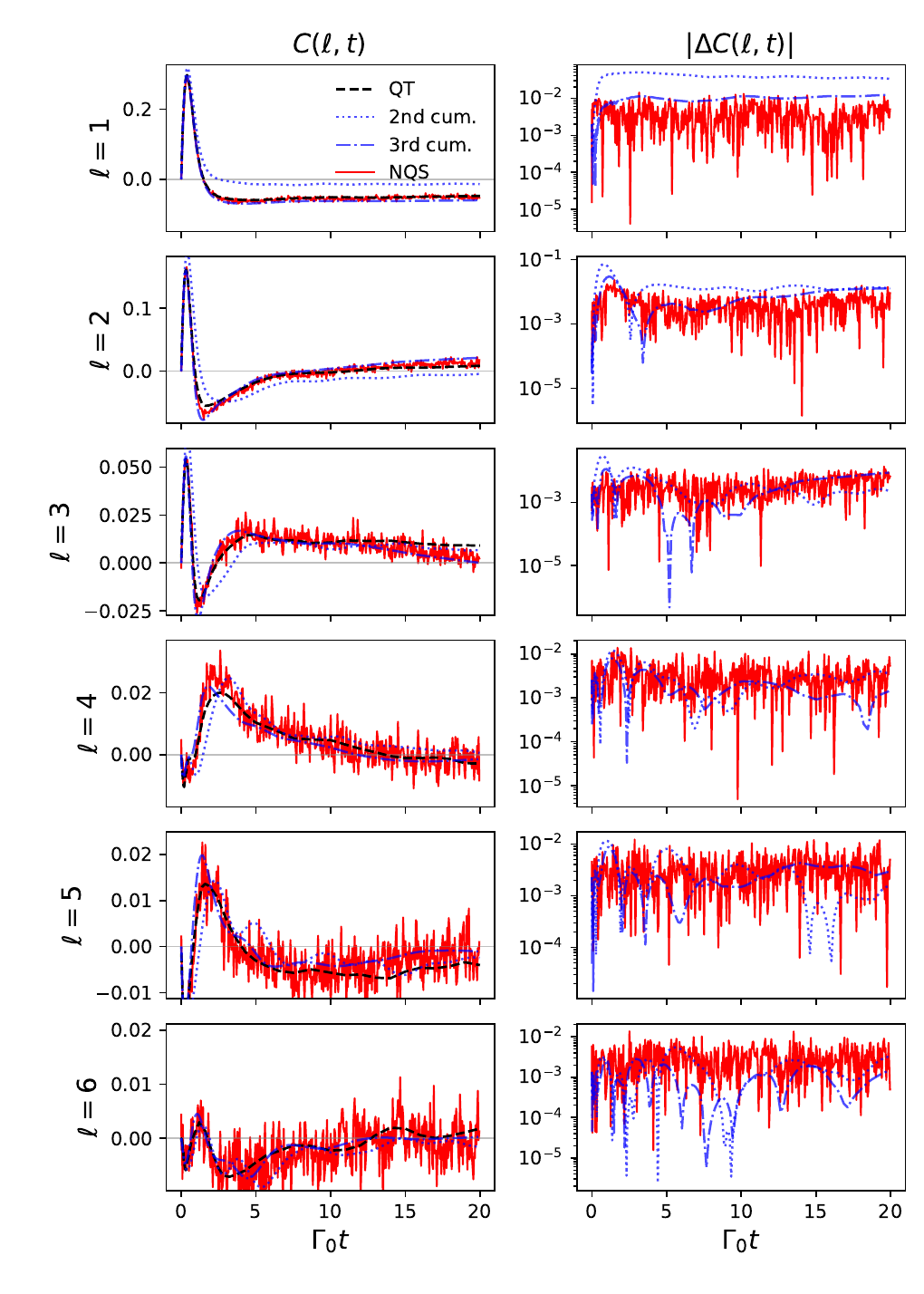}
    \caption{Comparison of connected correlations \eqref{eq:corr}, calculated by different approaches: quantum trajectories (QT), best NQS approximation from the main text ($f=12$, $d=4$, $k=4$) and cumulant expansions (2nd and 3rd order) for $L=16$ in 1D. Left column: Correlations $C(\ell,t)$ at different relative sites $\ell$. Right column: Absolute difference of correlations calculated by NQS or cumulant expansion with respect to the quantum trajectories result.}
    \label{fig:XXcorr}
\end{figure}

We test the ability of the neural quantum state (NQS) algorithm to capture the
spatial structure of the correlations generated during the dynamics.
To this end we evaluate the equal-time connected spin--spin correlator

\begin{equation}\label{eq:corr}
C(\ell,t)=
\langle \sigma^x_{\frac{L}{2}}\sigma^x_{\frac{L}{2} \; + \; \ell}\rangle(t)
-
\langle\sigma^x_{\frac{L}{2}}\rangle (t)\langle \sigma^x_{\frac{L}{2} \; + \;\ell}\rangle (t),
\end{equation}
which isolates two-body correlations by subtracting the disconnected
contribution.
In the numerical simulations the correlations are evaluated with respect
to the lattice site $L/2$, where $L$ denotes the system size and $\ell$ labels the relative lattice sites.
This choice provides a direct probe of how correlations spread across the
system during the time evolution.

Left column of \cref{fig:XXcorr} shows the time dependence of $C(\ell,t)$ for 1D dynamics presented in the main text, starting from the fully excited system that is evolved by Lindblad equation \eqref{eq:lindblad} for the parallel polarization. Correlators obtained from the best NQS POVM representation from the main text ($f=12$, $d=4$, $k=4$) are compared to the quantum trajectories result and the two different orders of cumulant expansion.  Convergence analysis with respect to different network sizes was performed on $L=10$ (not shown).  We find reasonably good agreement between the NQS and the QT result that could be further systematically improved upon increasing the network size. Fluctuations in the NQS result are due to the Monte Carlo sampling of the network to obtain the expectation values, see Supplementary Material for details.

The right column of \cref{fig:XXcorr} shows the absolute deviations of different approximations (NQS, 2nd and 3rd order cumulant expansion) to the quantum trajectories result. Absolute deviations are comparable to the deviations observed for the density of excited atoms, shown in \cref{fig:resnet1d}~(b), underlying that the neural network POVM representation is not fine-tuned to represent one observable better than the others.  
\newpage
\newpage
\newpage

\renewcommand{\thetable}{S\arabic{table}}
\renewcommand{\thefigure}{S\arabic{figure}}
\renewcommand{\theequation}{S\arabic{equation}}
\renewcommand{\thepage}{S\arabic{page}}

\renewcommand{\thesection}{S\arabic{section}}

\onecolumngrid

% \title{Iterative construction of conserved quantities in dissipative nearly integrable systems}
% \author{Iris Ul\v{c}akar}
% \affiliation{Jo\v{z}ef Stefan Institute, 1000 Ljubljana, Slovenia}
% \affiliation{University of Ljubljana, Faculty for physics and mathematics, 1000 Ljubljana, Slovenia}
% \author{Zala Lenar\v{c}i\v{c}}
% \affiliation{Jo\v{z}ef Stefan Institute, 1000 Ljubljana, Slovenia}
\setcounter{figure}{0}
\setcounter{equation}{0}
\setcounter{page}{0}

\newpage

\begin{center}
{\large \bf Supplemental Material:\\
Neural network modeling of many-body super- and sub-radiant dynamics}\\
\vspace{0.3cm}
Gianluca Lagnese$^{1}$, Laurin Brunner$^{2}$, Lorenzo Rossi$^{3}$, Darrick Chang$^{3,4}$, Markus Schmitt$^{5,6}$ and Zala Lenar\v ci\v c$^{1}$\\
\vspace{0.1cm}
$^1${\it Department of Theoretical Physics, J. Stefan Institute, SI-1000 Ljubljana, Slovenia} \\
$^2${\it Theoretical Physics III, Center for Electronic Correlations and Magnetism, Institute of Physics, University of Augsburg, D-86135 Augsburg, Germany} \\
$^3${\it ICFO–Institut de Ciencies Fotoniques, The Barcelona Institute of Science and Technology, 08860 Castelldefels (Barcelona), Spain} \\
$^4${\it ICREA–Instituci\'{o} Catalana de Recerca i Estudis Avan\c{c}ats, 08015 Barcelona, Spain
} \\
$^5${\it Forschungszentrum J\"ulich GmbH, Peter Gr\"unberg Institute, Quantum Control, 52425 J\"ulich, Germany
} \\
$^6${\it Faculty of Informatics and Data Science, University of Regensburg, D-93040 Regensburg, Germany} \\
\end{center}

In the Supplemental Material, we give more details on: (i) the cumulant expansion, (ii) the POVM representation of density matrix, and (iii) the TDVP modeling in the context of neural network approximation.

\vspace{0.6cm}

\twocolumngrid

\label{pagesupp}

\section{Cumulant Expansion}
For completeness, we report some details on the simulation technique based on the cumulant expansion (see Ref.~\cite{bigorda2023} for further details). This approach provides a systematic extension of mean-field theory. The construction proceeds as follows: (i) identifies the relevant operators---in the present case, products of Pauli operators $\sigma_i^{\alpha}$---up to a given order $n$; (ii) derives their equations of motion using the Heisenberg equation
\begin{equation}
    \frac{d \mathcal{O}}{dt} =  \cl{L}^{\dagger}(\mathcal{O}) \, ;
\end{equation}
(iii) imposes that connected correlators (cumulants) of order higher than $n$ vanish, thereby expressing $(n+1)$-point correlators in terms of lower-order ones and obtaining a closed set of equations. In our case, the Liouvillian $\cl{L}$ is given by \cref{eq:lindblad} of the main text.

The mean-field approximation is recovered at first order in the cumulant expansion. For the dynamics generated by \cref{eq:lindblad} and the initial state $\ket{\psi_0} = \prod_i \sigma_i^+ \ket{0}$, the only nontrivial observable is in this case $\braket{\sigma_i^z}$. The mean-field closure corresponds to imposing $\braket{\sigma_i^+ \sigma_j^-}_c = 0$, i.e.,
\begin{equation}
\braket{\sigma_i^+ \sigma_j^-} = \braket{\sigma_i^+}\braket{\sigma_j^-} \, .
\end{equation}
However, $\braket{\sigma_i^{\pm}} = 0$ for all times, leading to a trivial exponential decay.

The first nontrivial approximation is obtained at second order, where the relevant observables are $\braket{\sigma_i^z}$, $\braket{\sigma_i^+ \sigma_j^-}$, and $\braket{\sigma_i^z \sigma_j^z}$. The hierarchy is closed by setting third-order cumulants to zero \cite{kubo1962}, yielding
\begin{align}
\braket{\sigma_i^{\alpha} \sigma_j^{\beta} \sigma_l^{\gamma}} &=
\braket{\sigma_i^{\alpha}} \braket{\sigma_j^{\beta} \sigma_l^{\gamma}}
+ \braket{\sigma_j^{\beta}} \braket{\sigma_i^{\alpha} \sigma_l^{\gamma}}
+ \braket{\sigma_l^{\gamma}} \braket{\sigma_i^{\alpha} \sigma_j^{\beta}} \notag \\
&\quad - 2 \braket{\sigma_i^{\alpha}} \braket{\sigma_j^{\beta}} \braket{\sigma_l^{\gamma}} \, .
\end{align}

The equations for the second order cumulant expansion then read 
\begin{widetext}
\begin{align}
\frac{d}{dt} \braket{\sigma_i^z} &= -\Gamma_0 \braket{\sigma_i^z}
+ \sum_{n \neq i}
\Bigg[
\left( i J^{ni} - \frac{\Gamma^{ni}}{2} \right) \braket{\sigma^+_n \sigma^-_i}
 +
\left( - i J^{in} - \frac{\Gamma^{in}}{2} \right) \braket{\sigma^+_i \sigma_n^-}
\Bigg],\\[1em]
\frac{d}{ dt} \braket{\sigma_i^+ \sigma_j^-}
&= -\Gamma_0 \braket{\sigma_i^+ \sigma_j^-}
+ \frac{\Gamma^{ji}}{2}
\left(
4\, \braket{\sigma^z_i \sigma_j^z}
- \braket{\sigma^z_i} - \braket{\sigma^z_j}
\right) +  + i J^{ji} (\sigma^z_j - \sigma^z_i) + \notag\\
&+ \sum_{n \neq i,j}
\Bigg[
\left(
i J^{jn} + \frac{\Gamma^{jn}}{2}
\right)
\braket{\sigma^+_i\sigma^-_n}\,(2 \braket{\sigma^z_j} - 1)
+
\left(
- i J^{ni} + \frac{\Gamma^{ni}}{2}
\right)
\braket{\sigma^+_n \sigma^-_j} \, (2\braket{\sigma^z_i} - 1)
\Bigg],
\notag\\[1em]
\frac{d}{dt} \braket{\sigma_i^z \sigma^z_j}
&= -2\Gamma_0\, \braket{\sigma_i^z \sigma^z_j}
+ \sum_{n \neq i,j}
\Bigg[
\left(
i J^{nj} - \frac{\Gamma^{nj}}{2}
\right)
\braket{\sigma^z_i}\, \braket{\sigma^+_n \sigma^-_j}
+
\left(
- i J^{jn} - \frac{\Gamma^{jn}}{2}
\right)
\braket{\sigma^z_i}\, \braket{\sigma^+_j \sigma^-_n}
\notag\\
&\quad
+
\left(
i J^{ni} - \frac{\Gamma^{ni}}{2}
\right)
\braket{\sigma^z_j}\, \braket{\sigma^+_n \sigma^-_i}
+
\left(
- i J^{in} - \frac{\Gamma^{in}}{2}
\right)
\braket{\sigma_j}\, \braket{\sigma^+_i \sigma^-_n}
\Bigg].\notag
\end{align}
\end{widetext}

For the third order cumulant expansion the procedure is identical, with details and full equations reported in the appendix of Ref.~\cite{bigorda2023}. In a nutshell: (i) to the relevant operators from the second order expansion, the operators $\sigma_i^z\sigma_j^z \sigma_k^z$ and $\sigma_i^z\sigma_j^+ \sigma_k^-$ are added; (ii) the equations are written for each operator; (iii) the hierarchy is closed requiring that the fourth order cumulants are zero $\braket{ \sigma^{\alpha}_i \sigma^{\beta}_j\sigma^{\gamma}_k \sigma^{\delta}_l}_c=0$ \cite{kubo1962}.

\section{POVM Representation}

As described in the main text, the density matrix can be described by outcome probabilities $P(\mathbf a) = \text{tr}(M^\mathbf{a}\rho)$ of a positive operator-valued measure (POVM).
When the POVM is informationally complete (IC-POVM), the relation between probabilities and density matrix can be inverted
\begin{align}\label{eq:density_matrix_from_POVM}
    \rho=\sum_{\mathbf{a},\mathbf a'}P(\mathbf a)\left(T^{-1}\right)^{\mathbf a,\mathbf a'}M^{\mathbf a'}\, ,
\end{align}
with the overlap matrix $T^{\mathbf a,\mathbf a'}=\text{tr}\big(M^{\mathbf a}M^{\mathbf a'}\big)$.
For a composite Hilbert space $\mathcal H=(\mathcal H_{\text{local}})^{\otimes N}$, a many-body IC-POVM can be constructed via the tensor products of the measurement operators $M_{\text{loc}}^a$ of a local IC-POVM \cite{Carrasquilla2019}, $M^{\mathbf a}=M_{\text{loc}}^{a_1}\otimes\ldots\otimes M_{\text{loc}}^{a_N}$.
We choose the tetrahedral single-qubit POVM $M_\text{loc}^a = \frac{1}{2}|\psi_a\rangle\langle \psi_a|$, where the $|\psi_a\rangle$ span a tetrahedron on the Bloch sphere \cite{Carrasquilla2019,Reh2021}.
Thus, the density matrix is fully described when all $4^N$ outcome probabilities $P(\mathbf a)=\text{tr}\left(\rho M^{\mathbf a}\right)$ are known.
Inserting the expansion \eqref{eq:density_matrix_from_POVM} into the Lindblad equation yields a master equation for the probabilities
\begin{equation}
    \frac{d}{dt}P(\mathbf a)=\sum_{\mathbf a'}\mathscr L^{\mathbf a\mathbf a'}P(\mathbf a')\, ,
\end{equation}
where the operator $\mathscr L$ is derived from the Liouvillian $\mathcal L$ \cite{Reh2021}.
Similarly, expectation values of an observable $O$ can be calculated from the POVM probabilities by $\langle O \rangle = \text{tr}(O\rho) = \sum_\mathbf{a} P(\mathbf a) \Omega^\mathbf{a}$, where $\Omega^\mathbf{a} = \sum_{\mathbf{a}'}\left(T^{-1}\right)^{\mathbf a,\mathbf a'} \text{tr}\left(M^{\mathbf{a}'} O\right)$ is independent of the state of the system.

Due to the exponentially increasing Hilbert space size, expectation values --- including the components of the TDVP equation, $S_{k k^{\prime}}$ and $F_k$ --- need to be approximated using a Markov Chain Monte Carlo (MCMC) sampling method.
We build the Markov Chain by proposing a single site change in the POVM outcome $\mathbf{a} \to \mathbf{a}'$ and define the acceptance probability $r=\frac{P_\theta(\mathbf{a}')}{P_\theta(\mathbf{a})}$ using the neural network $P_\theta$.
More specifically, expectation values read
\begin{equation}
    \braket{\mathcal{O}} \approx \frac{1}{N_{\text{MC}}}\sum_{\mathbf{a} \sim P_\theta(\mathbf{a})} \Omega^\mathbf{a}~,
\end{equation}
where $a\sim P_\theta(\mathbf{a})$ signifies that the configurations $\mathbf{a}$ are sampled according to the distribution $P_\theta(\mathbf{a})$.
\vspace{0.2cm}

\section{Details of the time-dependent variational Monte Carlo}
For NQS, the stable integration of a TDVP equation requires careful regularization and the suited choice of a number of hyperparameters. 
Besides the established techniques of using a Heun integrator with adaptive time step based on the Fubini-Study metric with tolerance $\epsilon_{\mathrm{step}}$ and a signal-to-noise-ratio-based regularization of the TDVP equation with tolerance $\epsilon_{\mathrm{SNR}}$ \cite{Schmitt2020, jvmc, Medvidovic2023,Nys2024}, we introduce an adaptive variant of the truncation of the pseudo-inverse $S^+$: Instead of a fixed cutoff-parameter that determines the spectral truncation, we define a tolerance $\epsilon_{\mathrm{inv}}$ as a target accuracy limiting the residual of the solution of the TDVP equation,
\begin{align}
    \lVert S S^+F-F\rVert\overset{!}{<}\epsilon_{\mathrm{inv}}\ .
\end{align}
The truncation parameter $\tau$ of the pseudo-inverse $S^+\equiv S^+(\tau)=V\Lambda^+(\tau)V^\dagger$ is chosen adaptively as the maximal value, that yields the desired target accuracy. Here, $S=V\Lambda V^\dagger$ is the eigendecomposition with $\Lambda_{jj}\equiv\lambda_j$ and $\Lambda^+(\tau)$ is diagonal with $\Lambda_{jj}^{+}(\tau)=\left[\lambda_{j}\left(1+\left(\frac{\tau}{|\lambda_{j}/\lambda_{1}|}\right)^6\right)\right]^{-1}$, assuming descending ordering of the eigenvalues.
%
% We employ the techniques described in Refs.~\cite{Schmitt2020, jvmc, Medvidovic2023,Nys2024}, namely a Heun integrator with adaptive time step based on the Fubini-Study metric with tolerance $\epsilon_{\mathrm{step}}$, a soft-cutoff pseudo-inverse with truncation parameter $\epsilon_{\mathrm{inv}}$, and an signal-to-noise-ratio-based regularization with tolerance $\epsilon_{\mathrm{SNR}}$.
%

As mentioned above, we use another soft cutoff for the Monte Carlo estimation of components of the TDVP equations as introduced in Ref.~\cite{Schmitt2020}. More precisely, the strength term $F$ is rotated in the basis of $S$ as $\rho  = V^{\dagger}F$ and its signal-to-noise ratio $SNR(\rho)$ is evaluated. In addition to the previous regularization, $\Lambda_{jj}^+(\tau)$ is multiplied also by $\left(1+\left(\frac{\epsilon_{SNR}}{SNR(\rho)}\right)^6\right)^{-1}$. Such solution smoothly suppresses the components with a signal-to-noise ration smaller than $\epsilon_{SNR}$.

For our simulations, we found $\epsilon_{\mathrm{step}}\approx10^{-4}$, $\epsilon_{\mathrm{inv}}\approx10^{-6}$, and $\epsilon_{\mathrm{SNR}}=2$ to be suited choices for the regularization parameters while we draw of the order of $N_{\mathrm{MC}}\approx 10^5$ samples by Markov Chain Monte Carlo.
To ensure that the samples are sampled correctly from the desired distribution and have no auto-correlations, we do $5 \times L$ ($L$ being the system size) Markov Chain sweeps between two configurations and thermalize them by $8$ sweeps.

\end{document}